\documentclass[natbib]{svjour3}                     
\smartqed  
\usepackage{graphicx}

\newcommand{\commentpcf}[1]{}
\newcommand{\commenthans}[1]{}

\newcommand{\Vhc}{$V_{\mathrm{HC}}$ }
\newcommand{\kms}{\hbox{km s$^{-1}$}}

\newcommand{\nHI}{\hbox{$n({ \mathrm{ H^\circ} })$}}
\newcommand{\NHI}{\hbox{$N({ \mathrm{H^\circ}})$}}
\newcommand{\NaI}{\hbox{  Na$^\circ$} }
\newcommand{\DI}{\hbox{  D$^\circ$} }
\newcommand{\CaII}{\hbox{  Ca$^{+}$} }

\newcommand{\HeI}{\hbox{  He$^\circ$}}
\newcommand{\HI}{\hbox{  H$^\circ$}}
\newcommand{\nel}{\hbox{$n_{  \mathrm{e}}$}}
\newcommand{\npro}{\hbox{$n_{  \mathrm{p}}$}}
\newcommand{\cmtwo}{\hbox{cm$^{-2}$}}
\newcommand{\cc}{\hbox{cm$^{-3}$}}
\newcommand{\deeg}{\hbox{$^\circ$}}

\newcommand{\glon}{\hbox{$\ell$}}
\newcommand{\glat}{\hbox{$b$}}

\newcommand{\Beten}{\hbox{$^\mathrm{10}$Be}}
\newcommand{\Cltsix}{\hbox{$^\mathrm{36}$Cl}}
\newcommand{\Cfour}{\hbox{$^\mathrm{14} $C}}
\newcommand{\ebv}{\hbox{$E(B-V)$}}

\newcommand{\phiheI} {\hbox{$\phi_\mathrm{HeI}$}}
\journalname{SSrv}
\newcommand{\aap}{{Astron. Astrophys.}}

\newcommand{\apj}{{Astrophys. J.}}
\newcommand{\apjs}{{Astrophys. J. Suppl.}}
\newcommand{\apjl}{{Astrophys. J. Lett.}}
\newcommand{\grl}{{Geophys. Res. Lett.}}

\newcommand\aj{\ref@jnl{AJ}}%
\newcommand\actaa{\ref@jnl{Acta Astron.}}%
\newcommand\araa{{ARA\&A}}%
\newcommand\aarv{\ref@jnl{Astron. Astro. Rev.}}%
\newcommand\ao{\ref@jnl{Appl.~Opt.}}%
\newcommand\apss{\ref@jnl{Ap\&SS}}%
\newcommand\azh{\ref@jnl{AZh}}%
\newcommand\baas{\ref@jnl{BAAS}}%
\newcommand\caa{\ref@jnl{Chinese Astron. Astrophys.}}%
\newcommand\cjaa{\ref@jnl{Chinese J. Astron. Astrophys.}}%
\newcommand\icarus{\ref@jnl{Icarus}}%
\newcommand\jcap{\ref@jnl{J. Cosmology Astropart. Phys.}}%
\newcommand\jrasc{\ref@jnl{JRASC}}%
\newcommand\memras{\ref@jnl{MmRAS}}%
\newcommand\mnras{{MNRAS}}%
\newcommand\na{\ref@jnl{New A}}%
\newcommand\nar{\ref@jnl{New A Rev.}}%
\newcommand\pra{\ref@jnl{Phys.~Rev.~A}}%
\newcommand\prl{\ref@jnl{Phys.~Rev.~Lett.}}%
\newcommand\pasa{\ref@jnl{PASA}}%
\newcommand\pasp{\ref@jnl{PASP}}%
\newcommand\pasj{\ref@jnl{PASJ}}%
\newcommand\qjras{\ref@jnl{QJRAS}}%
\newcommand\rmxaa{\ref@jnl{Rev. Mexicana Astron. Astrofis.}}%
\newcommand\skytel{\ref@jnl{S\&T}}%
\newcommand\sovast{\ref@jnl{Soviet~Ast.}}%
\newcommand\ssr{{Space~Sci.~Rev.}}%
\newcommand\zap{\ref@jnl{ZAp}}%
\newcommand\nat{{Nature}}%
\newcommand\iaucirc{\ref@jnl{IAU~Circ.}}%
\newcommand\aplett{\ref@jnl{Astrophys.~Lett.}}%
\newcommand\apspr{\ref@jnl{Astrophys.~Space~Phys.~Res.}}%
\newcommand\bain{\ref@jnl{Bull.~Astron.~Inst.~Netherlands}}%
\newcommand\fcp{\ref@jnl{Fund.~Cosmic~Phys.}}%
\newcommand\gca{\ref@jnl{Geochim.~Cosmochim.~Acta}}%
\newcommand\jcp{\ref@jnl{J.~Chem.~Phys.}}%
\newcommand\jgr{{J.~Geophys.~Res.}}%
\newcommand\jqsrt{\ref@jnl{J.~Quant.~Spec.~Radiat.~Transf.}}%
\newcommand\memsai{\ref@jnl{Mem.~Soc.~Astron.~Italiana}}%
\newcommand\nphysa{\ref@jnl{Nucl.~Phys.~A}}%
\newcommand\physrep{\ref@jnl{Phys.~Rep.}}%
\newcommand\physscr{\ref@jnl{Phys.~Scr}}%
\newcommand\planss{\ref@jnl{Planet.~Space~Sci.}}%
\newcommand\procspie{\ref@jnl{Proc.~SPIE}}%

\begin{document}

\title{Time-variability in the Interstellar Boundary Conditions of the Heliosphere:
Effect of the Solar Journey on the Galactic Cosmic Ray Flux at Earth}

\titlerunning{Solar Journey}        

\author{Priscilla C. Frisch \and Hans-Reinhard Mueller
}

\authorrunning{Frisch and Mueller} 

\institute{Priscilla C. Frisch  \at
              University of Chicago, Chicago, IL\\
              \email{frisch@oddjob.uchicago.edu}           
           \and
           Hans-Reinhard Mueller \at
           Dartmouth College, Hanover, NH \\
     \email{hans.mueller@dartmouth.edu}
}

\date{Received: date / Accepted: date}

\maketitle

\begin{abstract}

During the solar journey through galactic space, variations in the
physical properties of the surrounding interstellar medium (ISM)
modify the heliosphere and modulate the flux of galactic cosmic
rays (GCR) at the surface of the Earth, with consequences for the
terrestrial record of cosmogenic radionuclides  One phenomenon that needs
studying is the effect on cosmogenic isotope production of
changing anomalous cosmic ray fluxes at Earth due to variable
interstellar ionizations. The possible range of interstellar ram pressures and ionization
levels in the low density solar environment
generate dramatically different possible heliosphere
configurations, with a wide range of particle fluxes of
interstellar neutrals, their secondary products, and GCRs
arriving at Earth. Simple models of the distribution and
densities of ISM in the downwind direction give cloud transition
timescales that can be directly compared with cosmogenic
radionuclide geologic records.  Both the interstellar data and
cosmogenic radionuclide data are consistent with two cloud transitions,
within the past 10,000 years and a second one 20,000--30,000 years ago,
with large and assumption-dependent uncertainties.
The geomagnetic timeline derived from cosmic ray fluxes
at Earth may require adjustment to account for the disappearance
of anomalous cosmic rays when the Sun is immersed in ionized gas.


\end{abstract}

\section{Introduction} \label{intro}

The Sun has traversed multiple interstellar clouds during it 220
Myr journey around the galactic center, including dense neutral
clouds, low density partially ionized interstellar matter (ISM)
such as now surrounds the heliosphere, and hot very tenuous
plasma.  Variations in the cosmic ray fluxes at the surface of the
Earth are strongly dependent on the geomagnetic field, the solar
magnetic activity cycle, the heliosphere, and the physical
properties of the circumheliospheric interstellar material. For
example, the modulation of the 1 AU cosmic ray flux by solar flare
mass ejections has long been known \citep{Gosling:1964forbush}.

The heliosphere acts
as a weather vane for the circumheliospheric ISM (CISM),
responding to the ionization, magnetic pressure, and
dynamic ram pressure \citep{Holzer:1989}.
The cosmic ray component at 1 AU varies with the properties
of the heliosphere modulation region.  The interpretation
of the geological record of cosmogenic isotopes relies on accurate
models of the cosmic ray spectra.  One factor that is not 
included in the interpretation of the geological record of
cosmogenic isotopes is that the cosmic ray spectrum incident on
the Earth consists of two components that behave differently as
the Sun travels through space. Galactic cosmic rays dominate at
high energies, $> 500 $ MeV, and are subject to heliospheric
modulation as the Sun travels through space. However a second
cosmic ray component at lower energies is formed inside of the
heliopause from interstellar neutrals that penetrate and are
ionized inside of the heliosphere, forming pickup ions.  These
are subsequently accelerated to form lower-energy anomalous cosmic
rays (ACRs) with a composition derived from neutral interstellar
atoms in the CISM \citep{Fisk+Ramaty_1974}. \commentpcf{ Any
scenario that connects features in the geomagnetic record with
interstellar cloud encounters will be based on assumptions about
the ISM.}  The local interstellar cosmic ray spectrum that
creates the geological radio-isotope record is thus composed of two
components that vary differently over time and space, the higher
energy galactic cosmic rays (GCRs) that are modulated by a
variable heliosphere, and the ACRs that also depend on the density
and fractional ionization of the surrounding interstellar cloud.

In this paper we present the overall picture of the ISM
characteristics that result from the motion of the Sun and
interstellar clouds through space.  Observations of interstellar
absorption lines towards nearest stars show that spatial
variations in velocity, temperature, and ionization of the
circumheliospheric ISM create temporal variations in the
heliosphere boundary conditions.  These then cause temporal
variations in the spectrum and fluxes of cosmic rays at Earth. We
also draw possible connections between interstellar cloud
transitions and the geological radio isotope record.


\section{Contemporary and Paleointerstellar Circumheliospheric ISM} \label{sec:ism}

\subsection{Environment, Dynamics, Structure, and  Magnetic Field in
  Nearby ISM}  \label{sec:clouds}

Interstellar clouds establish the boundary conditions of the heliosphere,
and the heliosphere is the primary modulation region for the galactic cosmic
ray flux at Earth.  We therefore briefly review the properties
of nearby ISM that sets the future and paleointerstellar CISM.

The Sun is traveling through a region 
with very low mean densities that extends 70--200
pc from the Sun (Figure 1).
\footnote{In Fig. 1, the cumulative
dust distributions in and beyond the Local Bubble boundaries are
evaluated from the color excess \ebv\ and astrometric data in the
Hipparcos catalog (\cite{Perryman:1997}), after first ignoring
stars with variability as shown by the Hipparcos index I$ > 0$.
The ISM close to Sco-Cen stars, and associated with the 18.5\deeg\
tilt of Gould's Belt, are within the range of $|$Z$| < 50$ pc.}
Stellar winds and supernovae
in star formation regions bordering the Local Bubble both sculpted the
Local Bubble and contributed to the cosmic ray flux at Earth.
Hot low density plasma is widespread in the Local Bubble
cavity , $T\sim 10^6$ K, $n < 0.005$ \cc\
\citep[][F09]{Frischetal:2009}.  
Within $\sim 15$ pc, a cluster of
local interstellar clouds is flowing past the Sun.  
Clouds with densities of $\approx 0.3$ \cc\ and $\approx 10^3$ \cc\ have been
identified in this flow.  Despite the large difference in scale sizes between the
heliosphere, with a distance to the upwind heliopause of $\sim 150$ AU (0.0007 pc), 
and the Local Bubble, the heliosphere traces the solar environment
that is set by the Local Bubble interior, in particular the interstellar radiation field and 
magnetic field.
The Local Bubble environment of the Sun affects the past, present,
and future heliosphere boundary conditions.

The heliosphere varies over geologically short timescales
due to the velocities of the Sun and wispy local interstellar clouds through space.
Fig. 1 shows the vector motion of the Sun through the interior of the
Local Bubble, based on the solar apex motion through the local
quasi-inertial frame known as the local standard of rest
\citep[LSR, we use a solar velocity of $ 18.0 \pm 0.9 $
\kms, towards \glon,\glat $ = 47.9^\circ \pm 3.0^\circ,~
23.8^\circ \pm 2.0^\circ$, based on results in
][]{BinneyDehnen:2010}.  


The heliospheric boundary conditions are set by the CISM.  Interstellar
neutral gas and dust are replenished in the heliosphere over
timescales of $\sim 30$ years due to the relative 26 \kms\ motion
of the Sun and CISM.
Self-consistent photoionization models of the
surrounding ISM  provide
a good match to data on the ISM inside and adjacent to the heliosphere
\citep[][S08]{SlavinFrisch:2008}.  
\footnote{Note there are different velcity models for the Local Interstellar
Cloud \citep[LIC,][]{Lallementetal:1995,FGW:2002,RLIV:2008vel}; we assume that
the ISM inside of the heliosphere is part of the LIC seen towards
Sirius, 2.7 pc, and $\epsilon$ CMa.  Recent IBEX observations are
consistent with the conclusion that the Sun is in the LIC cloud
seen towards Sirius (see Frisch and McComas, this volume).}  
These models predict densities and temperatures for the CISM of
\nHI=0.19 \cc, \nel=0.065 \cc, \npro=0.055 \cc, and
T=6,300 K (from Model 26 in SF08).  The fractional ionizations of
H and He are 22\% and 39\%, respectively.  Constraints on
the models include 
the density and temperature of interstellar He
inside of the heliosphere, the local ISM towards $\epsilon$ CMa (and
Sirius), pickup ion data giving cloud neutrality, and an
interstellar radiation field that includes 
ultraviolet, soft X-ray, and
extreme ultraviolet emissions, including contributions from a conductive boundary on the
local interstellar cloud.
Regardless of whether the Local Bubble plasma contains hot plasma,
the models and data indicate that the CISM is warm, partially ionized and low density
(see Model 42 in SF08).  

The ISM within $\sim 30 pc$  has been
shocked, since the gas-phase abundances are characteristic of the
pattern expected from the destruction of refractory grains in
50--100 \kms\ shock fronts \citep{Frischetal:1999}.  
An association of the local ISM with an evolved superbubble
shell is thus suggested.  The abundances for the refractory
elements Mg, Si, and Fe are larger by factors of 4--8 than
abundances of a cold cloud, and C is overabundant.
Comparisons of interstellar gas-phase Mg, Fe, and Si abundances with
solar abundances show that the local grains are olivine silicates,
that can be efficiently aligned by magnetic fields \citep{Frischetal:2011araa}.  

The properties of the circumheliosphere ISM have varied rapidly over time.  LSR velocities of
the Sun and most nearby interstellar clouds are near 18 \kms\ (or
$\approx 18$ pc Myrs$^{-1}$); however the cluster of local
interstellar clouds (CLIC, within 30 pc) is part of a general flow of
ISM past the Sun so that many relative Sun-cloud velocities are larger
than 14 \kms\ (Table 1).  The LSR vector motion of the CLIC is --16.8
\kms\ arriving from the upwind direction \glon$\sim 335^\circ$,
\glat$\sim -7^\circ$ (Figure 1).  This direction is towards the center
of the Loop I (North Polar Spur) supernova remnant \citep[at
\glon,\glat=320\deeg,5\deeg, ][]{Heiles:1998whence},
and makes an angle of $70^\circ \pm 35^\circ$
with the direction of the local interstellar magnetic field
\citep{Frischetal:2010ismf}.  These relative Sun-CLIC velocities and
the configuration of local ISM (Figure 1) indicate that the Sun has
recently emerged from the hot Local Bubble plasma and entered the
CLIC.

Cloud velocity is a  key interstellar variable affecting the heliosphere boundary
conditions, because the interstellar ram
pressure varies as the square of the Sun-cloud velocity. The CLIC is a
decelerating flow.
From upwind to downwind, cloud velocities relative to the Sun
($V_{\mathrm{HC}}$) are --28.4 \kms\ towards 36 Oph, 26.3 \kms\ in the
inner heliosphere \citep[according to ][]{Moebiusetal:2004}, and 23.4 \kms\ towards $\chi^1$ Ori (Table 1).  For
otherwise similar clouds, these velocity differences lead to a 50\%
difference in the ram pressures of the ISM on the heliosphere over
timescales of less than $\approx 4000$ years.  Over longer timescales
of $10^5$ years, the heliocentric cloud velocities and cloud lengths
in Table 1 suggest variations of a factor of $\sim 4$ in interstellar ram
pressures.

The distribution of the low column density local clouds,
\NHI$<10^{18.5}$ \cmtwo, is patchy.  The volume density of neutral
gas in nearby clouds is known only for the CISM. If all local
clouds have this same density, \nHI=0.2 \cc, then $\sim 35$\% of
the sightlines to stars within 10 pc are filled with warm low
density ISM.  The clouds have a mean thickness $0.9 \pm 0.3$ pc, and
the Sun would cross these clouds with a mean crossing time of
$\sim 47,000$ years.

\begin{figure*}[t!]
\begin{center}
  \includegraphics[width=0.45\textwidth]{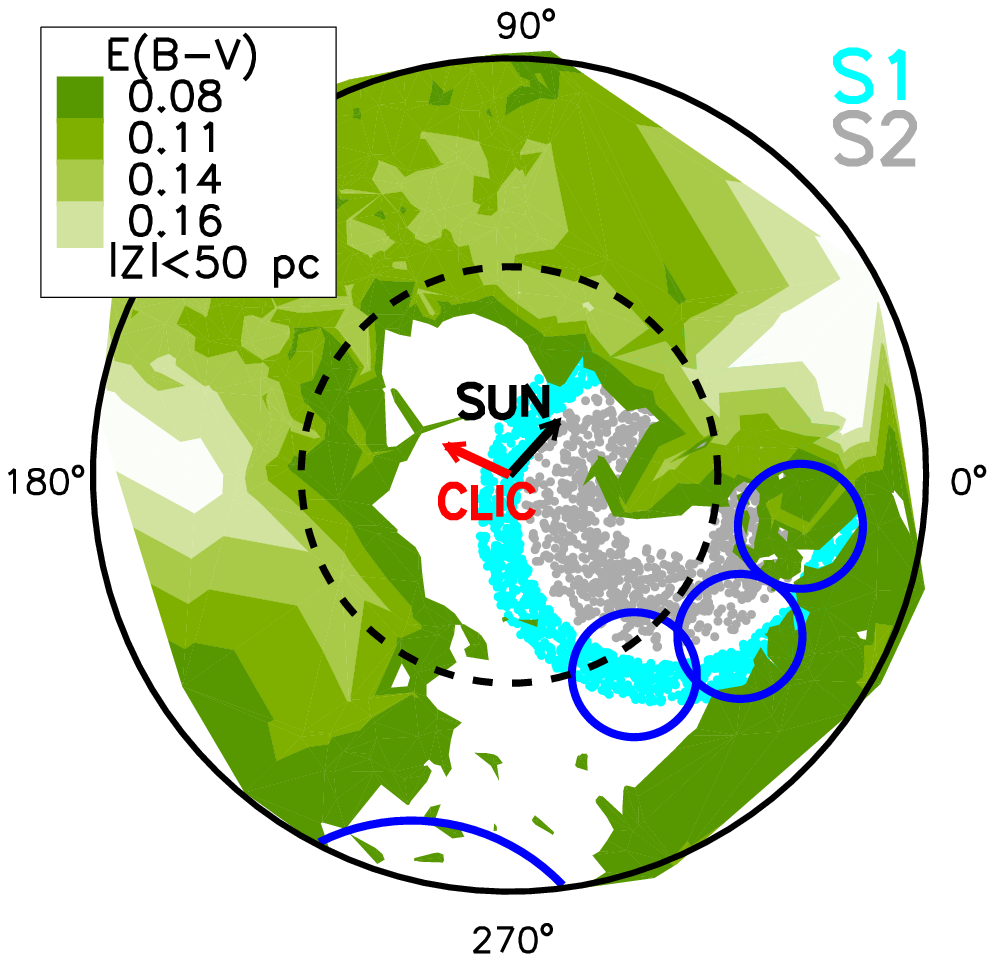}
  \includegraphics[width=0.45\textwidth]{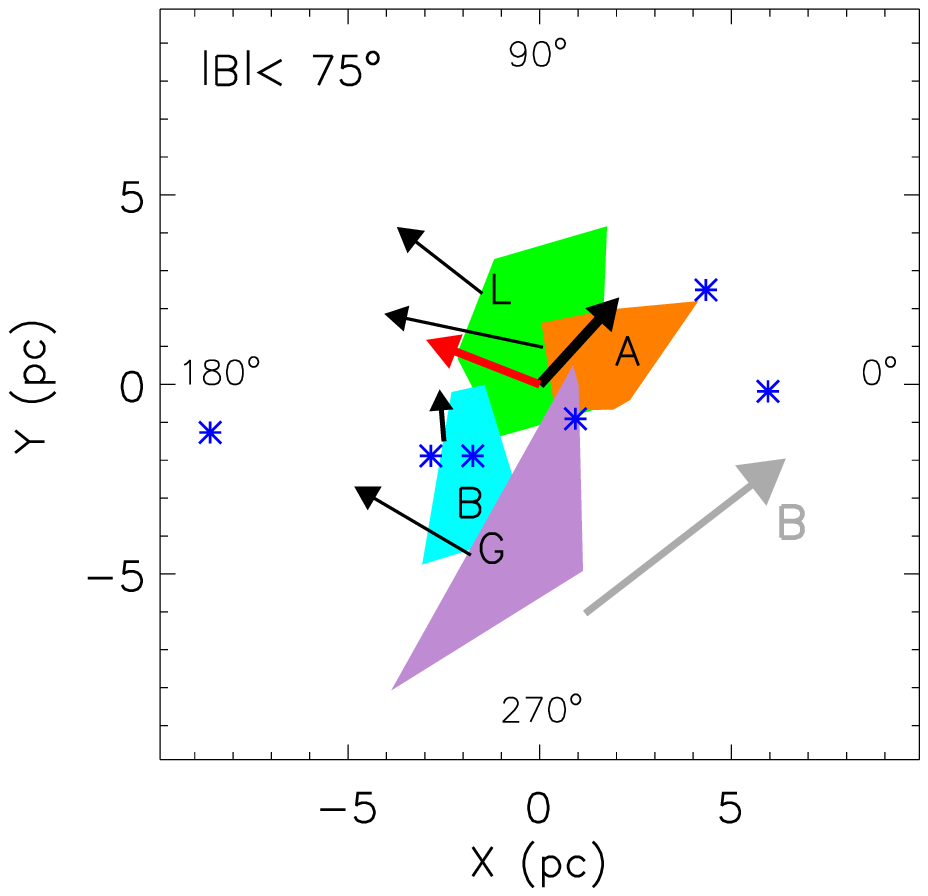}
\end{center}
\caption{Left: The distribution of interstellar dust within 200 pc of the Sun
and 50 pc of the galactic plane, according to the cumulative
amount of ISM traced by color excess E(B-V).
The solar and CLIC motions through the LSR are given, respectively, by
the thick black and red arrows (in both figures).  The intersection of the S1 (cyan) and S2
(gray) magnetic shells of Loop I with the galactic plane are shown
based on spherical shells from \citet{Wolleben:2007}.  The \ebv\
contour levels of 0.08, 0.11, 0.14, 0.16 mag correspond to
$N$(\HI+$H_2$) column densities of 4.46e20, 6.56e20, 7.87e20, and
9.18e20 \cmtwo, for $N$(\HI+$H_2$)/\ebv=5.8e21 atoms \cmtwo\ mag
(\cite{BohlinSD:1978ebv}).  The large blue circles show the three
subgroups of the Sco-Cen Association; the arc centered near \glon$\sim
260^\circ$ shows the approximate nearside of the Gum Nebula.  Right:
The locations of the nearest stars in Table 1 are plotted in the galactic
plane, together with the interstellar magnetic field direction
\citep[gray arrow, from ][]{Frischetal:2010ismf}, and interstellar
clouds within 5 pc of the Sun and $75^\circ$ of the galactic
plane.  The clouds that are plotted include
the LIC and the Blue clouds (green, blue,
from RL08), and the G-cloud and Apex clouds \citep[lavender and orange,
from ][] {Frisch:2003}.  The clouds are labeled, and LSR velocities are
shown.  }
\label{fig:lb}
\end{figure*}

Magnetic fields permeate the region of the CLIC, and create
asymmetries in the heliosphere configuration.  
The Sun appears to be located in or near the rim of the "S1" 
magnetic shell associated with Loop I \citep[e.g.][]{Wolleben:2007,Frisch:2010s1}.  
Magnetically aligned
interstellar dust grains near the Sun create a birefringent medium with
lower optical opacities for directions parallel to the interstellar
magnetic field (ISMF), and therefore polarize starlight.
The direction of ISMF
over the nearest 40 pc has been found from fits to the position angles
of polarized starlight, giving a direction towards \glon, \glat$ ~ = ~
38^\circ, ~ 23^\circ$, with uncertainties of $\sim \pm 35^\circ$
\citep{Frischetal:2010ismf}.  The $\sim 70^\circ $ angle between the
magnetic field and upwind LSR CLIC direction is consistent with an ISMF
compressed in an expanding superbubble shell.  The ISMF direction
indicated by the Interstellar Boundary Explorer (IBEX) Ribbon arc
center, \glon,\glat$=33^\circ, 55^\circ$ \citep[][this
volume]{FrischMcComas:2010} is consistent with the polarization
direction to within uncertainties.  A more distant measure of the ISMF
is provided by fits to the Faraday rotation and dispersion measures
for four pulsars, 150--300 pc away in the third galactic quadrant,
which give a ISMF direction similar to the polarization value, and
indicate a field strength of $\sim 3.3$ $\mu$G \citep{Salvati:2010}.

Not all nearby clouds are warm and diffuse.  Towards the constellation
of Leo, and within 12 pc, a tiny dense cold filamentary interstellar
cloud with thickness $<0.4$ pc has been identified in \NaI\ absorption
\citep{MeyerLauroesch:2006}.  
This cloud is similar to the tiny scale atomic structures (TSAS) 
that are observed throughout the ISM, with typical sizes 30 AU, volume
densities $\sim 10^3$ \cc, column densities $10^{18} - 10^{19}$
\cmtwo, and thermal pressures $P/k=nT=10^4 - 10^6$ \cc K 
\citep{Stanimirovic:2009}.  
Once the solar motion is removed from cloud velocities (Figure 1, right), 
the small Blue cloud is also seen to form in
a region where the LIC and G-cloud are colliding.
Both the Leo cloud and the Blue cloud (e.g. HD 80007) coincide with the ring of
tiny dense cold clouds identified in \HI\ 21-cm by \citep{Haud:2010}.
The CLIC is a decelerating flow of ISM, so that a scenario where
the upwind G-cloud and downwind LIC are converging is consistent
with the velocity data.
Evidently tiny cold clouds can and have formed very close to the Sun
through cloud collisions, so that the recent paleoheliosphere may
well have included intervals where galactic cosmic ray modulation
was minimal (see next section).  

Other possible types of nearby ISM include conductive,
evaporative, boundaries between warm diffuse gas and hot plasma
between the clouds.  Unsaturated outflows of ISM from the clouds lead
to sharp gradients in the ISM density, velocity, and temperature over
small spatial scales. The outflow velocity can be up to 20 \kms, with
temperature variations of an order of magnitude, over distances $<0.5$
pc (SF08).

\subsection{Configuration of Local ISM}

Our goal is to explore the connection between the paleoCISM and the
paleomagnetic record on Earth.  This connection is possible
because of the heliosphere response to variations in the circumheliospheric
ISM, and the cosmic ray flux at Earth is connected both to the
heliosphere and neutrality of surrounding ISM. Optimally
we need a model of the properties of nearby clouds, including
cloud shapes, density, ionization, velocity, and
homogeneity.  Most of these data are unavailable, so we rely
on simple assumptions for the cloud configuration and density.

Discussions of the effect of the ISM on the heliosphere
will be restricted to clouds that extend to within 5 pc of the Sun.
For a typical Sun-cloud velocity of 17 \kms, this would
correspond to a look-back time of $\sim 300,000$ years.
The most complete data set available for determining cloud
configurations are the \HI\ and \DI\ data, which are free of
uncertainties regarding elemental abundances \citep{RLII,Woodetal:2005}.  Column densities
for \HI\ or \DI\ can be transformed to distance scales
characteristic of the neutral gas for some assumed volume density
for \HI. 
We will assume that nearby clouds have the same neutral
density as the LIC, \nHI=0.2 \cc.

Clouds near the Sun are typically identified through
parsing absorption line velocity data into different clouds
by assuming rigid-body motion through space for each cloud.
A number of studies provide data on cloud velocities, but the
most complete study is that of
\citet[][RL08]{RLIV:2008vel}.  However, one of the conclusions of this study,
that the LIC does not surround the heliosphere, may change
with better measurements of the velocity of interstellar \HeI\ inside of the
heliosphere (Frisch and McComas, this volume).
Distinct interstellar "clouds" have been found within 5 pc in a number
of different studies \citep[e.g. see review ][]{Frischetal:2011araa}.
We adopt the LIC and Blue cloud (B)  properties from
RL08, and the G cloud (G) and Apex cloud (A) from \citep{Frisch:2003}; additional
column density data is provided by \citet{Woodetal:2005} and \citet{RLII}.  
The configuration of these clouds, within 60$^\circ$ of the galactic plane,
is shown in Figure 1, right. The LSR velocity vectors for the clouds 
are shown, together with those of the CLIC and Sun.  
For the purposes of estimating cloud
thicknesses from \CaII, the conversion factor of $N$(\CaII)/$N$(HI)$=1.5 \times  10^{-8}$
is adopted from $\alpha$ Aql and $\eta$ UMa data.

Estimates of the times it took (or will take) for the Sun to cross
interstellar clouds observed towards several nearby stars are listed
in Table 1, for the simplest assumption that the time is given by
$T=L/V_\mathrm{HC}$, for $L=$\NHI/\nHI, heliocentric velocity
$V_\mathrm{HC}$, and \nHI$=0.2$ \cc.  For stars near the upwind or
downwind directions, this assumption should give realistic estimates
as long as the cloud is homogeneous and neutrals and ions are well
mixed.  Obviously irregular cloud shapes, or cloud motions from
unmeasured non-radial velocities, particularly for sightlines parallel
to cloud surfaces or for inhomogenous clouds, can affect these basic
estimates.  Also listed in Table 1 are the angles between each star
and the upwind directions of the vector CLIC, and vector LIC (from
\HeI), velocities through the LSR.  Whether clouds in the local ISM
are filamentary or blobs affects the time inferred for the Sun to
cross the cloud, particularly for stars making a large angle with the
upwind direction where radial velocities are small.

 \begin{table}[bt!]
\caption{Crossing times for Clouds Close to Sun $^{(1)}$} \label{tab:1}
\begin{tabular}{lcccccc}
\hline\noalign{\smallskip}
Star  & $\glon$, $\glat$,Dist,Cld & $V_{\mathrm{HC}}$, $V_{\mathrm{LSR}}$ &
\NHI & $L$, Cloud &  $\theta_{\mathrm{CLIC}}, \phi_{\mathrm{HeI}}$ &  $T$, Crossing
 \\
 &    &    &  &  Thick.  & LSR, HC   &  Time  \\
      & (deg,deg, pc)  &    (\kms)        &  (\cmtwo)        & (pc)         &  (deg)                & (years)  \\
\noalign{\smallskip}\hline\noalign{\smallskip}
 $\alpha$ Cen  & 316, --1, 1.3, G  &  --18.0,  --18.6  & 17.61 &  0.66  & 20\deeg, 130\deeg  &  (35,900) \\  	
 $\alpha$ CMa & 227, --9, 2.7, L  &  19.6,  2.2  & 17.60  &  0.65  & 106\deeg, 42\deeg  &  $ 32,200$  \\		
 $\alpha$ CMa & 227, --9, 2.7, B  &  13.7, --3.7  &17.40 &  0.41  & 106\deeg, 42\deeg  &  $ 29,000$  \\		
 $\alpha$ CMi & 214, 13, 3.5 & 24.0, 10.4 & 17.81 & 1.86 & 122\deeg, 41\deeg  & 42,600  \\			
 $\alpha$ CMi & 214, 13, 3.5, L & 20.5, 6.5 & 18.08 & 1.38 & 122\deeg, 41\deeg  & 92,900 \\
  70 Oph & 30, 12, 5.1, G & --26.2, --9.3 & 17.77   &  0.95 &  $58^\circ,~26^\circ$  &  (670,000) \\ 		
  70 Oph & 30, 12, 5.1, A & --32.4, --15.5 & 17.46   &  0.47 &  $58^\circ,~26^\circ$  & (530,000) \\ 
 $\alpha$ Aql & 48, --9, 5.1, G &  --18.1, -0.3 & 17.91 & 1.3 & $72^\circ,~50^\circ$  & (1.1e6) \\			
 $\alpha$ Aql & 48, --9, 5.1, A &  --26.9, --10.7 & 17.47 & 0.48 & $72^\circ,~50^\circ$  & (670,000) \\
 $\chi^1$ Ori &  188, --3, 9, L  &  22.3, 27.4 & 17.93  &  1.38  & 146\deeg, 13\deeg  &  60,500  \\			
 $\alpha$ Aur &  163, 4, 13, L   &  21.8, 15.3  & 18.24 &  2.82  & 171\deeg, 28\deeg  &  126,000 \\    		
 36 Oph       & 358, 7, 6, G &  --28.2, --16.7  & 17.85 &  1.15  & 26\deeg, 170\deeg  & (39,800)  \\    
\noalign{\smallskip}\hline

\noalign{\smallskip}
\end{tabular} \\
$^{(1)}$  Column 1 gives the star number in Figure 1,
and star name.  
The galactic coordinates, distance, and cloud to which the component is attributed, are listed in Column 2.
G and A refer to the G-cloud and Apex cloud \citep[from][]{Frisch:2003}, and L and B refer to the
LIC and Blue clouds \citep[from ][]{RLIV:2008vel}.  
Column 3 gives the observed
velocities of the interstellar absorption components in the solar
(e.g., \cite{RLIV:2008vel}) and LSR inertial systems
Column 4 is the log of the cloud column density, from \cite{Woodetal:2005}
or \cite{Hebrardetal:1999}.
Column 5 is the cloud thickness, \NHI/\nHI\
calculated for volume density \nHI$ = 0.2$ \cc. 
Column 6 gives the angles $\theta_{\mathrm{CLIC}}$
and $\phi_{\mathrm{HeI}}$, which are the angles between the
star and the LSR upwind CLIC direction (\glon,~\glat$~=~335^\circ,-7^\circ$),
and the heliocentric (HC) downwind \HeI\ vector (\glon,~\glat$~=~184^\circ,-15^\circ$), respectively.
Column 7 gives a nominal crossing time for the cloud, 
calculated from the cloud thickness and $  V_{\mathrm{HC}}$. The times inside
of parentheses are in the future.  \\ 

\end{table}

The times in Table 1 are estimates only, since accurate dating of
solar transitions of cloud boundaries requires data on the volume
densities that are not generally available.  If the clouds are in
thermal pressure equilibrium, the cloud temperature variations of a
factor of two or more \citep{RLIII} suggest there are also variations
in cloud densities, which can not be evaluated without information on
cloud ionization.  In addition, the structure of local ISM is resolved
at very low spatial resolution, since current data only sample the sky
at a spacing of $\sim 1$ star per 260 square-degrees (e.g. RL08).

\subsection{Possible Historical Changes in the Circumheliospheric ISM  } \label{sec:history}

Stars located close to the downwind directions of cloud velocities
will give the best estimates for entry times into the cloud, for
simple geometries.  For the LIC, downwind corresponds to small values
for \phiheI.  Estimates of the time when the Sun entered in the LIC
will be based on interstellar data towards four stars, two near the
downwind direction of the LIC ($\chi^1$ Ori and $\alpha$ Aur), and two
at more oblique angles to the LIC vector ($\alpha$ CMi, and $\alpha$
CMa).  The Sun is assumed to be in the LIC.  Because of the 
uncertainties in the ionization, volume density, magnetic pressure,
an homogeneity of these clouds, the timescales below are quite uncertain.
However they do show the plausibility of an interstellar effect on the
cosmogenic isotope record.

The star $\chi^1$ Ori, at 9 pc, is viewed through the heliosphere tail
(\phiheI$=13^\circ$, Table 1).  The observed \Vhc$=22$ \kms\ component
corresponds to the LIC velocity.  For the above premise that
$L=$\NHI/\nHI, the Sun would have entered the LIC $\sim 60,500$ years
ago.  The relatively high turbulence of the LIC absorption lines
towards $\chi^1$ Ori \citep[e.g. ][]{RLIV:2008vel}, however, suggests that two
clouds with similar velocities could blend in velocity to form this
component.  If so, the Sun would have entered the LIC more recently.
A more distant star in the downwind direction is $\alpha$ Aur
(\phiheI$=28^\circ$).  The higher LIC \HI\ column densities
suggest, instead, that the the Sun entered the ISM 126,000 years
ago. The star $\alpha$ CMi is much closer to the Sun, but it is offset
41\deeg\ from the downwind LIC direction (\phiheI=41\deeg).  The LIC
component gives an entry time into the LIC of 92,900 years ago.

The closest downwind star is Sirius ($\alpha$ CMa, 2.7 pc), with
two clouds in front of it.  Sirius is $106^\circ$ from the
downwind direction.  The Sun would have entered the LIC component
32,200 years ago, and the Blue cloud $\sim 29,000$ years
before that.  

An alternate LIC encounter time 
follows from the assumption that the LIC is filamentary, with the 
filament oriented perpendicular to the LSR velocity and roughly
parallel to the magnetic field direction.
When distant diffuse ISM is spatially resolved, filamentary structures
are generally observed for neutral and ionized gas.  Magnetic pressure
creates filamentary ISM towards Loop I at distances of $\sim 90$ pc,
so the evident relation between Loop I and nearby ISM 
suggest the LIC could also be filamentary.  
\citet{Frisch:1994sci} assumed
a filamentary structure for the LIC, with the direction of
the LSR cloud velocity perpendicular to the filament, 
and found the Sun may have entered the CISM gas anytime within the past
2,000--8,000 years, or even more recently.

For the comparisons with the geologic radio isotope record
in Section 4, and since the detailed structure of the downwind
gas is not known, we adopt several time intervals as candidates
for a solar transition between interstellar clouds.  The primary
downwind clouds of interest are the LIC, which is most likely
presently surrounding the Sun, and the Blue cloud (Figure 1,
right).  These transitions are: (1) Sometime within the past 8,000
years, as indicated by the $\alpha$ CMa sightline together with an
assumed filamentary cloud structure. (2) Sometime within $10,000 -
32,000$ years ago.  (3) Before that the Sun was in the Blue cloud,
with an entry time of $\sim 61,000$ years ago from the $\alpha$
CMa timescales.  The Blue cloud
should be more ionized than the LIC because it is
less shielded from the primary source of local H-ionization,
$\epsilon$ CMa.  This suggests that the Sun traveled through a
cloud with higher ionization levels than the LIC, before it
entered the LIC.  This interpretation is also consistent with the
local ionization gradient found by \citet{Wolffetal:1999}.  (4)
Any of the surfaces of these clouds may have a conductive
interface. However the clumpiness of the local ISM, and the
apparent deceleration evident in the CLIC velocities, suggest that
the clumps of gas that form the CLIC have a cohesive origin and
are close to each other.  Transition times across 0.5 pc cloud
interfaces, where the relative Sun-cloud velocities may be larger
by $\sim 20$ \kms, may be 10,000 years, or less, depending on the
angle of the ISMF that inhibits conduction.  (5) No tiny dense
clouds have been directly detected nearby in the downwind direction,
however the Blue cloud coincides with a ring of tiny cold dense
clouds (Section \ref{sec:clouds}).
Typical sizes for these clouds, as small as 30 AU, indicate
that such clouds would sweep past the heliosphere over
timescales of a decade.

\section{The Heliosphere for Different Interstellar Environments} \label{sec:heliosphere}

The heliosphere results from a balance between solar wind and
interstellar pressure. This balance sets the size of the heliosphere
and determines the particle distributions and magnetic field
configurations throughout, which in turn determine how galactic cosmic
rays are being modulated during their passage through the heliosphere
on their way to Earth.

\begin{figure}[t!h]
\begin{center}
\includegraphics[width=0.49\textwidth]{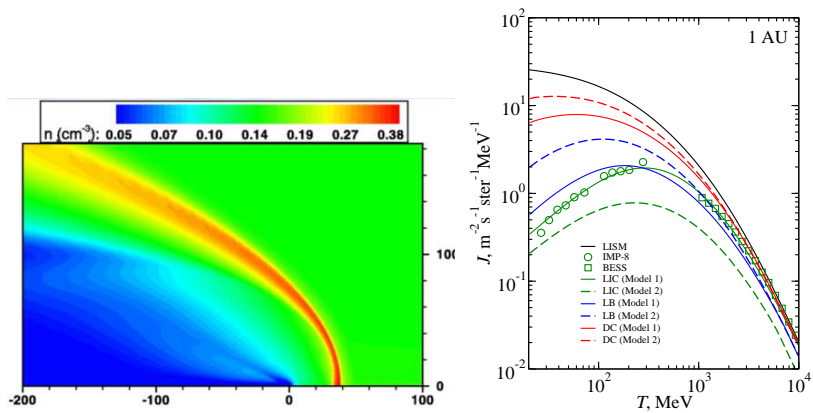}
\includegraphics[width=0.37\textwidth]{fig2b.eps}
\end{center}
\caption{{\em Left.} 
The density distribution for the heliosphere when immersed in
a moderately dense cloud, \nHI$ \sim 15$ \cc, with a temperature T=3,000 K,
ion density of 0.2 \cc, and moving at the LIC velocity \citep[Model 17 in ][]{Muelleretal:2006}.
The heliopause is 34 AU from the Sun.
{\em Right.} Spectra of galactic cosmic
rays at 1 AU for three different interstellar clouds surrounding the
heliosphere (ACRs are not included).  LIC stands for the contemporary CISM; LB is the Local
Bubble interior modeled as a $1.2 \times 10^6$ K fully ionized plasma
with density 0.005 \cc; and DC is a dense cloud with density 10 \cc,
T=200 K, and relative velocity of 25 \kms.  Models 1 and 2
utilize different cosmic ray modulation models \citep[figure
from][]{FlorinskiZank:2006jos}.  } \label{fig:3}
\end{figure}

The inner boundary conditions of the heliosphere consist of the solar
wind, imprinted with the 22 year magnetic activity cycle of the Sun,
as a ``point source'' inflow. The time variability of the solar wind
affects the large-scale heliosphere, including the heliosheath
regions immediately inside and outside of the heliopause.
The distance of the termination shock in the nose direction
varies by about 10 AU with the solar cycle phase, and is closer during
solar wind minima when ram pressures are lower
\citep{ZankMueller:2003}. The reversals of the solar magnetic polarity
every $\sim 11$ years between successive solar minima propagate to the
inner heliosheath regions, and create bands of magnetic polarity that
are swept around the flanks of the heliosphere with the subsonic solar
wind \citep[e.g.,][]{Pogorelovetal:2009}. The outer boundary
conditions of the heliosphere are dominated by the total pressure of
the surrounding cloud, including magnetic pressure, the ram pressure
of excluded ions, and a large fraction of the ram pressure of the
neutrals which participate through charge-exchange with ions outside
the heliopause; the pressure interior of the heliopause is modified by
charge exchange as well.

Configurations of the global heliosphere have been modeled for a
range of surrounding interstellar cloud properties. In one study
\citep{MuellerWoodman:2008}, the parameter space around the
current interstellar cloud is sampled to establish empirical
relations for locations of heliopause, termination shock, and bow
shock as a function of interstellar boundary values. This study
confirms the overall connection of the size of the heliosphere
with the pressure balance argument. The interstellar ionization
ratio weakly influences the downstream termination shock distance
as well as the typical termination shock strength. In relative
terms, the termination shock shape does not change in this
particular region of parameter space.  \citet{Mueller:2006}
explore a wide range of possible interstellar environments,
including cold dense clouds, hot tenuous, completely ionized
clouds like that assumed for the Local Bubble, and systems on
galactic paths that result in fast relative Sun-ISM velocities. In
particular the relative velocity has a decisive influence on how
much interstellar neutral material reaches the inner heliosphere
(filtration); the larger the relative velocity, the less
filtration is occurring. If the Sun encounters a completely
ionized ISM, such as the Local Bubble, GCR particle fluxes at Earth are lower
than currently. With neutrals absent, none of the pressure balance
modifications take place, and secondary particles like anomalous
cosmic rays (ACR) do not exist. Lastly, if the heliosphere is
embedded in a dense cloud, the heliosphere is small in size, and
particle fluxes at Earth rise substantially. Depending on the
interstellar ram pressure, the heliosphere can easily get small
enough for the Earth orbit to be partly in the inner heliosheath
region.

\citet{YeghikyanFahr:2006jos} consider the passage of very dense
molecular clouds over the heliosphere. The resulting heliospheres are
very small, so that the orbit of Earth takes it through regions of
interstellar gas outside the heliopause. The expected neutral fluxes
at Earth are so high that changes for the ozone layer and other
climate-related effects are to be expected.  Similarly, when a
supernova shock front washes over the heliosphere
\citep{Muelleretal:2009ssr}, the heliosphere can be equally small
(Figure \ref{fig:3}a), with direct access of supernova material to the
Earth atmosphere.  In both cases of extremely small heliospheres, the
access of GCR to Earth is greatly enhanced. In a calculation by
\citet{FlorinskiAxford:2003}, where a cold dense cloud results in a
moderate size heliosphere of 23 AU, a two-fold flux of GCR, and a
ten-fold flux of ACR is reported. Figure \ref{fig:3}b illustrates this
with calculated spectra for the Local Bubble case and for a dense
cloud case, comparing them to the GCR fluxes of the contemporary
heliosphere.

\section{Comparing past cloud transitions with the terrestrial radioisotope record}\label{sec:geomagnetic}

Spallation of cosmic-rays in the terrestrial atmosphere creates the
radioisotopes \Beten, \Cltsix, and \Cfour\ that are used to date the
repository geological archive.  Both the geomagnetic and heliosphere
magnetic fields modulate the incident cosmic-ray flux
\citep[e.g. ][and the Beer and McCracken articles in this
volume]{McCracken:2004ice}, including the ACR component. The well
known anticorrelation between cosmic ray fluxes at 1 AU and the solar
magnetic activity cycle (see articles by Leske and Mewaldt in this
volume) suggests that ISM-driven variations in the heliosphere may
also be significant.  Earlier interstellar explanations for the peaks
in the concentration of \Beten\ in the ice core record include the
reduced modulation of cosmic rays by a heliosphere that has been
compressed by a passing supernova shock \citep[][and see
Fig. 2]{SonettJokipii:1987}, and higher GCR fluxes incident on the
heliosphere because the Sun traveled through a magnetic flux tube in
the ISM \citep{Frisch:1997}.  High GCR fluxes have also been linked to
climate cooling \citep[e.g.,][]{KirkbyCarslaw:2006jos}.

The paleomagnetic field that is derived from geological data such as
magnetic remenance should, in principle, agree with the paleomagnetic
field that has been derived from the radioisotope \Beten\ and \Cltsix\
ice core data, and the \Cfour\ tree-ring data.  Reconstructing the geomagnetic
field from cosmogenic isotope data requires an assumption about the
flux and spectra of cosmic rays, and the production and dispersal of
radioisotopes.  \citet[][M05]{MuschelerBeer:2005} compared the
paleomagnetic field determined by these different techniques, and
found
several discrepancies that are not explained.
Given the mystery of these anomalies, we postulate that temporal
variations in the properties of the ISM shaping the
heliosphere led to variations in the total cosmic-ray spectrum at
the Earth.  These variations have been overlooked as a factor in the geomagnetic
time-line developed from cosmogenic isotopes, but need to
taken into consideration.

The anomalies in the reconstructed paleomagnetic record mentioned
in the \citet{MuschelerBeer:2005} paper, and that are
discussed here are: 1. The discrepancy in the timing of the maximum
geomagnetic field dipole strengths (the ``virtual axis dipole
moment'', VADM) during the late Holocene traced by the \Beten\ versus
\Cfour\ records (see Fig. 6 in M05).  2. The interval of 18,000--34,000 years ago,
where both \Beten\ and \Cltsix\ VADM reconstructions show increased
modulation compared to the remanence records, and the \Cfour\ data are poorly
understood.  3.  An anomaly 48,000 years ago where both \Beten\
and \Cltsix\ showed increased modulation compared to the VADM record,
and an anomaly 58,000 years ago where only \Cltsix\ shows increased
modulation.

Cosmic ray fluxes at Earth vary with the size and properties of the
heliospheric modulation region.  A second simple change in the CISM
properties, that would \emph{necessarily} affect cosmic ray fluxes at
the Earth, would be spatial variations in the ionization levels of the
interstellar gas.  At 1 AU the cosmic-ray fluxes below $\sim 200$
MeV/nucleon are dominated by anomalous cosmic rays (ACRs).  For
instance, the ACR oxygen intensity of 10 MeV/nucleon is an order of
magnitude larger than for GCR oxygen \citep{LeskeCummings:2008}.  ACRs
form from neutral interstellar atoms that charge-exchange with the
solar wind, creating pickup ions that are accelerated to cosmic ray
energies.  The ACRs will vanish if the surrounding ISM is ionized
(with the exception of a very minor contribution arising from
dust-generated pickup ions).  These spatial variations translate to
temporal variations in the 1 AU cosmic ray fluxes because of the
relative Sun-cloud motion.

Models of radioisotope production include the GCR sources, and cosmic
rays from energetic solar flares.  However, they do not distinguish
ACRs as a source with a distinct origin and spectrum, and one that is likely
to vary as the Sun encounters different clouds.  Radioisotope
production has been calculated for solar cosmic rays at similar
energies as ACRs \citep{WebberMcCracken:2007}, but not for the ACR
spectrum itself.  ACRs are trapped in the radiation belts of Earth
\citep{Mewaldt:1998acrrevewbelt}.

The primary question then becomes: How would the \Cfour, \Beten,
and \Cltsix\ production change as the ACR component vanishes?
Three possible effects may contribute. The first is that \Cfour\
is formed by thermalized neutrons near the top of the atmosphere,
and the long storage time ($>0.1$ year) of ACRs trapped in the
radiation belts of Earth \citep{Mewaldt:1998acrrevewbelt} may
multiply the production of \Cfour\ because of the long exposure
times of nitrogen and carbon compounds in the atmosphere compared
to the direct production of cosmogenic isotopes formed by
spallation.  An opposite effect of ACRs on cosmogenic isotope
production is suggested by the yield functions of \Beten\ and
\Cfour\ as a function of the energy of the incident cosmic ray
proton, which are shown in Figure 6 in \citet{Usoskin:2008}. The
production of \Beten\ is relatively more efficient at ACR energies
then the production of \Cfour, based on comparisons of the ratios
of production at $\sim 200$ MeV (ACR energies) and $\sim 20$ GeV
(GCR energies). These ratios for \Beten\ and \Cfour\ are $\sim
0.03$ and $\sim 0.001,$ respectively.  These yield functions would
then suggest that the production of \Cfour\ is less sensitive to
the ACR component of cosmic-ray fluxes than is the production of
\Beten.  A third wild-card possibility is that rapid variations of
the geomagnetic field may reduce coupling between the ACRs and the
radiation belts, so that ACRs have the same access to the
atmosphere as GCRs, reducing any effect of storage of ACRs in the
radiation belts.  In the absence of a detailed understanding of
the effect of ACRs on cosmogenic isotope production, the
discussions below linking cloud transitions to the geological
radioisotope data are highly speculative.


\subsection{Late Holocene Discrepancies and Anomalous Cosmic Rays}

The VADM paleomagnetic field reconstructed from \Beten\ records peaks
at 2,000 years BP, which is one millennium after the peak determined
from \Cfour\ at 3,000 years BP (Fig. 6 in M05).  At the CHISM
velocity, 3,000 years corresponds to a distance of 0.08 pc, which is a
plausible value for the distance to the cloud surface in the $\alpha$
CMa direction.  
The 2,000-3,000 time interval is within the uncertainties of the solar entry into
the LIC as inferred from the $\alpha$ CMa sightline by \citet{Frisch:1994sci},
based on a filamentary structure for the LIC.  A LIC filamentary structure
would be consistent with Figure 1, right, only if the density structure of
the ISM moving at the LIC velocity is inhomogeneous.  

Since VADM variations anticorrelate with radioisotope
fluxes, the \Cfour\ fluxes decreased a millennium before the \Beten\
fluxes decreased in the Holocene.  \citet{MuschelerBeer:2005}
postulated that this discrepancy was due to the carbon cycle, because
the \Beten\ and archeomagnetic field determinations generally agree.

If the discrepancy is due, instead, to variations in the ISM, we suggest that
ACRs may be the culprit.
The three possible influences of ACRs on the cosmogenic isotope
record leads to three different explanations for 
the anomaly.  Prior to entering the CISM, the Sun may have traversed a
fully ionized cloud interface with no neutrals, no pickup ions or ACRs,
with a larger relative Sun-cloud velocity leading to higher interstellar ram pressures (see
Fig. 2 in SF08), and with reduced mass-loading of the solar wind by
pickup ions.    
(1) If the enhanced yield of \Beten\ production at low
energies is important, the absence of ACRs at 1 AU would increase
\Cfour\ relative to \Beten, which is not seen.  (2) An opposite result
follows if the long exposure times for \Cfour\ formation from ACRs in
the radiation belts is important, in which case an ionized ambient ISM
would preferentially reduce \Cfour\ levels because of the concordant
lack of \Cfour\ enhancement from the multiplying effect of the storage
of ACRs in the magnetosphere.  Once the Sun enters the main part of
the cloud, which is $\sim 75$\% neutral, the low energy ACR component
is restored and the balance between the synchronization of the \Beten,
\Cltsix, and \Cfour\ formation was restored along with the ACRs.  (3)
For the third alternative, if disruption of the terrestial magnetic
field affects the \Cfour\ production only, \Beten\ would stay coupled
to the geomagnetic field while the timescale for \Cfour\ dating would
be disrupted.  The tight coupling between \Beten\ and the geomagnetic
field suggests that the overall heliospheric modulation of the GCR
fluxes did not change, so that the discrepancy could be due 
to the effect that the disrupted geomagnetic field has on the flux
of ACRs into the atmosphere.

\subsection{Cloud Crossings}

 During the period 18,000--34,000 years ago, both the \Beten\ and
\Cltsix\ VADM reconstructions deviate from the 
archeomagnetic remanence records. The increased VADM reconstructed from \Beten\
and \Cltsix\ suggests reduced fluxes below that explained by the
archeomagnetic record, or extra heliosphere modulation of GCRs. \citet{MuschelerBeer:2005}
suggested that possible changes in levels of solar magnetic activity
may be responsible for this difference.  We speculate that this
interval instead corresponds to an encounter between the Sun and
slower clumpy ISM beyond the CISM in the downwind direction that forms
the second clouds observed towards $\alpha$ CMa (the Blue cloud, Figure 1 right) 
and towards $\alpha$ CMi \citep[the Aur cloud in ][]{RLIV:2008vel}, both
within 3.5 pc.  Cloud velocities alone affect the GCR modulation
because the heliosphere dimensions increase as the ram pressure of the
ISM decreases.  The decreased \Beten\ and \Cltsix\ fluxes in this
interval would then suggest a larger heliosphere modulation region,
that still contains a solar wind mass-loaded with pickup ions from interstellar
neutrals.  Cloud-Sun relative velocities of 26 \kms\ today, versus
21--22 \kms\ for $\chi^1$ Ori and $\alpha$ Aur, and 14 \kms\ for the
Blue cloud towards $\alpha$ CMa (Table 1), suggest that cloud ram pressure
could have been a factor of $\sim 2-3$ lower in the past than today,
giving a larger heliosphere and larger GCR modulation region that are
consistent with the decreased fluxes found by M05.

Additional discrepancies, about 46,000--48,000 and 58,000 years ago, are seen
between the radio isotope paleomagnetic and remanence paleomagnetic
records.  
If the Sun entered the neutral portion of the LIC $\sim 60,500$ years ago, as indicated by
$^1 \chi$ Ori located in the downwind direction (Table 1), the VADM
discrepancy suggests 
that increased cosmic ray modulation in the LIC, or decreased ACR production,
could have produced the discrepancy. 
A slightly higher density, \nHI$\sim 0.25$ \cc, would match the solar entry 
into the LIC to 48,000 years ago discrepancy.

\section{Future:  Improving Comparisons between Local ISM
Structure and Paleomagnetic Records}

Comparisons between the structure and kinematics of interstellar
clouds near the Sun and the paleomagnetic records deduced from
radio isotopes such as \Beten, \Cltsix, and \Cfour, are based on
scanty knowledge of details of the three-dimensional spatial
distribution and configuration of nearby interstellar clouds.
Volume densities of both neutrals and electrons in the nearest
interstellar clouds are required to improve maps of the ISM
distribution.  Some insight could be gleaned from photoionization
models, as done for the CISM; however high-sensitivity ultraviolet
data are required for realistic models. 

The most significant open question is whether ACR variations
are sufficient to require consideration as a separate component
that contributes to the production of the cosmogenic isotopes,
and that varies separately from the GCR component as the heliosphere
is modulated by the ISM.  One aspect of
answering such questions will be to expand calculations of the
fluxes of galactic and anomalous cosmic rays at the Earth beyond
the limited number of heliosphere configurations now available.
More study of the roles of
ACRs versus higher GCRs as source populations of
the radio isotopes are needed, since differences between
the \Cfour, \Beten, \Cltsix\ records are mandated if ACRs are a
factor in \Cfour\ production rates. A better understanding
of the physical properties and configuration of ISM close
to the Sun is also required for accurate comparisons between
the timelines of the cosmogenic isotope and heliosphere
boundary conditions.
Nevertheless, we conclude that in principle the
attenuation of GCRs by an ISM-modulated heliosphere,
and perhaps the presence or absence of ACRs,
are capable of accounting for 
differences between the paleomagnetic record determined from cosmogenic
isotopes versus remnance data.

\begin{acknowledgements}
PCF thanks the International Space Sciences Institute in Bern,
Switzerland, for hosting a stimulating meeting on the relation between
cosmic ray fluxes and the terrestrial radio isotope record.  This
research has been supported in part by NASA grants NNX09AH50G and
NNX08AJ33G to the University of Chicago, and by the IBEX mission as a
part of NASA's Explorer Program.  PCF would like to thank Ken
McCracken and Jurg Beer for helpful discussions.

\end{acknowledgements}


\end{document}